\documentclass[twocolumn,pra,aps,showpacs,10pt]{revtex4-1} 
\usepackage{amsfonts}
\usepackage{amsmath}
\usepackage{amssymb}
\usepackage{color}
\usepackage{float}
\usepackage{hyperref}
\usepackage{textcomp}
\usepackage{subfigure}
\usepackage[rightcaption]{sidecap}
\usepackage{graphicx}
\begin{document}
\title{Multiple electromagnetically induced transparency, slow and fast light in hybrid optomechanics}
\author{M. Javed Akram$^{1}$}
\email{mjakram@qau.edu.pk}
\author{Fazal Ghafoor$^{2}$}
\email{rishteen@yahoo.com}
\author{Farhan Saif$^{1}$}
\email{farhan.saif@fulbrightmail.org}
\affiliation{$^1$Department of Electronics, Quaid-i-Azam University, 45320 \ Islamabad, Pakistan. \\ $^2$Department of Physics, COMSATS Institute of Information Technology, Islamabad, Pakistan}
\begin{abstract}
We theoretically investigate the phenomenon of electromagnetically induced transparency (EIT) of a weak probe field in hybrid optomechanics with a single three-level ($\Lambda$-type) atomic system. We report that, in the presence of optomechanical coupling and two transition coupling parameters of three-level atom (TLA), there occurs three distinct multiple EIT windows in the probe absorption spectrum. Moreover, the switching of multiple windows into double and single EIT windows can be obtained by suitably tuning the system parameters. Furthermore, the probe transmission spectrum have been studied. Based on our analytical and numerical work, we explain the occurrence of slow and fast light (superluminal) regimes, and enhancement of superluminal behaviour in the probe field transmission. This work demonstrates great potential in multi-channel waveguide, fiber optics and classical communication networks, multi-channel quantum information processing, real quality imaging, cloaking devices and delay lines \& optical buffers for telecommunication.
\end{abstract}
\date{\today}
\maketitle

\section{Introduction}
Electromagnetically induced transparency (EIT) \cite{Harris,Fleischhauer,Lukin,Agarwal,Akram,AH,Weis,Zubairy2} is a quantum interference phenomenon that takes place when two laser fields excite resonantly and two different transitions share a common state. Since its experimental realization \cite{Harris2,Harris3}, EIT has made valuable contributions in the development of optical sciences because of its promising applications in multidisciplinary fields \cite{Yang,Lukin2,Hakuta,Akram2,Wang1}. For instance, EIT has appeared as fundamental ingredient for the real applications like group velocity control \cite{Scully,Akram3,Ghafoor}, nonlinear susceptibility modulation \cite{Boyd}, quantum information processing \cite{Fleischhauer,AH}, quantum metrology \cite{AH,Akram}, dark-state polariton \cite{Akram4}, induced photon-photon interaction in cold atomic gases \cite{Fleischhauer}, Rydberg atomic-state detection \cite{Adams,Saffman}, cavity linewidth narrowing \cite{Xiao} and laser frequency stabilization \cite{Adams2}.

On the other hand, the studies on the EIT have been extended to multi-level atomic systems interacting with multiple laser fields \cite{CT,Knight,Wang,Wang2}. It has been reported that multiple EIT windows occur in the probe absorption spectrum in the multi-level atomic systems \cite{Yang2,Zubairy}, whose absorption profiles of a weak probe field are different from that of three-level $\Lambda$-system \cite{Fleischhauer}. Up to now, multiple EIT windows have been reported in various multi-level atomic systems, such as $V$-type \cite{Ying}, $Y$-type \cite{Safari}, $N$-type \cite{Ham} and $K$-type \cite{Sun} atomic systems. Unlike single EIT phenomenon, multiple EIT windows allow transmission of the probe field at multiple different frequencies simultaneously \cite{Huang}. Such multiple EIT phenomenon could be useful for promising applications in multi-channel optical communication, waveguides for optical signal processing and multichannel quantum information processing.

Recently, significant theoretical and experimental efforts \cite{AKM,Meystre} have been made to utilize radiation pressure forces in order to probe and to manipulate micro/nano-scale mechanical objects, which have paved the way for a number of new types of optomechanical systems. Amongst these, guided wave nanostructures in which large gradients in the optical intensity are manifest, have been shown to possess extremely large radiation pressure effects \cite{Painter,Marquardt,Lipson}. Nevertheless, the recent demonstration of strongly interacting photonic and phononic resonances in a quasi one-dimensional (1D) optomechanical crystal \cite{Painter} has shown that it may be possible to coherently control phonons, photons, and their interactions on an integrated, chipscale platform. On the other hand, the waveguide optomechanical interface possess a unique combination of high sensitivity, broad bandwidth, high quality single-crystal materials, and high $Q_{m}$, and has potential to allow measurement of quantum motion of nanobeam resonances \cite{Barclay}. It has been shown that the pump-probe response for the optomechanical (hybrid) systems shares the properties of paradigmatic $\Lambda$-type (multi-level) system, and displays EIT \cite{Agarwal, Akram} and double EIT phenomena \cite{Akram,Yang,Huang}.

More recently, Huang and Tsang \cite{Huang}, have shown multiple EIT-windows with $N$ membranes in optomechanics. Our present work is based on hybrid quantum optomechanics with three-level $\Lambda$-type atomic (TLA) system. In this paper, we report that: (i) Multiple (three) tunable EIT windows exist in the hybrid system containing single TLA. (ii) The presence of TLA in the system uniquely  provides us coherent control to switch from multiple to double and single EIT windows in a single experimental setup. (iii) We study the probe field transmission, and explain the occurence of slow and fast light regimes via multiple interferences. (iv) In addition, we explain the enhancement of superluminal behaviour of probe field in the hybrid system by controlling the system parameters. (v) The parametric values of the hybrid system used in our study are realizable in present-day laboratory experiments. Our analytical and numerical results obtained for these experimental parametric values show very good agreement.
\section{The Model Formulation}\label{sec2}
We consider a realistic hybrid cavity QED-optomechanical system, where a single cavity field mode is coupled both to a mechanical resonator as well as interacts with the $\Lambda$-type three-level atom (TLA) between the transitions $\vert 1 \rangle \leftrightarrow \vert 2 \rangle$. A strong laser field of frequency $\omega_l$, and a weak probe field of frequency $\omega_p$, simultaneously drive the optical cavity of length $L$. Moreover, a classical laser field with frequency $\nu_{c}$ induces the atomic transition $\vert 2 \rangle \leftrightarrow \vert 3 \rangle$ with coupling parameter $g_{1}$, which serves as a control field for the atom. Thus, the system has the usual nonlinear coupling of optomechanics $g_{o}$.  and two transition coupling parameters $g_{1}$ and $g_{2}$ associated with the $\Lambda$-type TLA.
\begin{figure}[ht]
\includegraphics[width=0.48\textwidth]{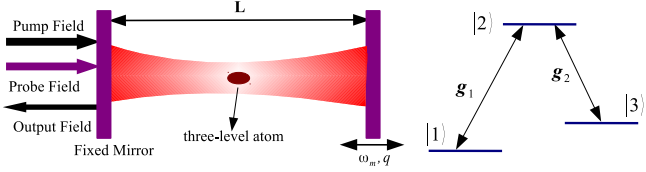}
\caption{The schematic representation of our hybrid system: A strong driving field of frequency, $\omega_l$, and a weak probe field of frequency, $\omega_p$, simultaneously drive the optical cavity of length $L$ containing a single three-level atom (TLA). The cavity field with frequency $\omega_c$ is coupled to a moving mirror with optomechanical coupling $g_{o}$, as well as interacts with the atom between the transition $\vert 1 \rangle \leftrightarrow \vert 2 \rangle$ (which acts as probe field for the atom) with coupling parameter $g_{2}$; whereas a classical laser field with frequency $\nu_{c}$ induces the atomic transition $\vert 2 \rangle \leftrightarrow \vert 3 \rangle$ with coupling parameter $g_{1}$, which serves as a control
field for the atom.} \label{cavity}
\end{figure}
Hence, in the rotating reference frame at the frequency $\omega_l$ of the strong driving field, the combined Hamiltonian of the system can be written as \cite{Meystre2}:
\begin{eqnarray}\label{Ham}
H_T &=& [\dfrac{p^2}{2m} + \dfrac{1}{2} m \omega_m^2 q^2] + \hbar \Delta_c c^{\dagger}c + \hbar \sum_{i=1}^{3} \Delta_{i} \sigma_{ii} \nonumber \\ &+&
\hbar (g_{1} c \sigma_{21} +g_{2}e^{-i\nu t}\sigma_{23} + H.c.)  - \hbar g_{o}c^{\dagger}cq \nonumber \\ &+& i\hbar\Omega_l(c^\dagger - c) +
i\hbar(\varepsilon_p e^{-i\delta t}c^\dagger - \varepsilon_p^* e^{i\delta t}c),\end{eqnarray}
where, $\Delta_c=\omega_c-\omega_l$, $\Delta_{i}=\omega_{i}-\omega_l$ ($i=1,2,3$) and $\delta=\omega_p-\omega_l$ are the detunings of the cavity field frequency, transition frequencies of the atom and the probe field frequency respectively, with respect to driving field frequency $\omega_l$. Here, the first term in Eq.~(\ref{Ham}), gives the free Hamiltonian of the moving mirror. The parameters $q$ and $p$, represent the position and momentum operators of the mirror with the vibration frequency $\omega_m$ and the mass $m$. The second term describes the Hamiltonian of the single cavity field mode with the creation (annihilation) operator $c^\dagger$ ($c$). The third term shows the free Hamiltonian of the TLA \cite{Meystre,Zubairy2}, where $\omega_{i}$ (i=1,2,3) is the transition frequency of atomic state $\vert i \rangle$ of the TLA, and $\sigma_{ii}$ are the raising and lowering operators of the TLA with associated transition frequency $\omega_i$. The fourth and fifth terms describe the interaction Hamiltonian; where fourth term shows the interaction between the cavity field and the moving mirror with optomechanical coupling $g_{o}=\frac{\omega_c}{L}\sqrt{h/m\omega_m}$. The fifth term denotes the atom-field interaction, where $g_1 = -\mu \sqrt{\omega_1/2V \epsilon_0}$ is the coupling constant. Here $\mu$ is the electric-dipole between the two levels ($\vert 1 \rangle$ and $\vert 2 \rangle$), $V$ is the cavity mode volume, and $\epsilon_0$ represents the permittivity of vacuum. Moreover, the classical laser field with frequency $\nu_{c}$ induces a transition $\vert 2 \rangle \leftrightarrow \vert 3 \rangle $, which is known as a control
field for the TLA with coupling parameter $g_{2}$. Finally, the last two terms corresponds to the classical fields, pump and probe lasers, with frequencies $\omega_l$ and $\omega_p$ respectively. Moreover, $\Omega_l$ and $\varepsilon_p$ are related to the laser power $P$ by $|\Omega_l|=\sqrt{2\kappa P_{l}/ \hbar \omega_l}$ and $\varepsilon_p=\sqrt{2\kappa P_{p}/ \hbar \omega_p}$ respectively, with $\vert\Omega_l\vert >> \vert\varepsilon_p\vert$.

We adopt the interaction picture for the atomic levels and work in the sideband resolved regime i.e. $\omega_m \gg \kappa$ \cite{Agarwal}. We obtain steady-state solutions of the system operators and study the output spectrum by using the mean field approximation (factorization assumption) \cite{Agarwal,Akram}, viz, $\langle qc\rangle =\langle q\rangle \langle c\rangle $. Thus, assuming that the field and the atomic system correlation functions factorize, the equations of motion for the expectation values of the atom-field system in the cavity are then given as \cite{Agarwal2}:
\begin{subequations}\label{meq}
\begin{align}
\dfrac{d\langle p \rangle}{dt}&=-m\omega_m^2 \langle q \rangle - \gamma_m \langle p \rangle + g_{o}\langle c^\dagger \rangle \langle c \rangle, \\
\dfrac{d\langle q \rangle}{dt}&=\dfrac{\langle p \rangle}{m}, \\
\dfrac{d\langle c \rangle}{dt}&=-(\kappa+i\Delta_c) \langle c  \rangle +i g_{o}  \langle c \rangle \langle q  \rangle -ig_{1} \langle \tilde{a} \rangle \nonumber\\ &+\Omega_l + \varepsilon_p e^{-i\delta t}, \\
\dfrac{d\langle \tilde{a} \rangle}{dt}&=-(\gamma_{1}+i\Delta_{1}) \langle \tilde{a} \rangle  - i g_{1} \langle c  \rangle - i g_{2} \langle \tilde{b} \rangle, \\
\dfrac{d\langle \tilde{b} \rangle}{dt}&=-(\gamma_{2}+i\Delta_{2}) \langle \tilde{b} \rangle  - i g_{2} \langle \tilde{a} \rangle,
\end{align}
\end{subequations}
where, $\tilde{a}=\sigma_{21}$, $\tilde{b}=e^{i(\nu_{c} -\omega_{l}) t} \sigma_{13}$, $\Delta_{1}=\omega_{21}-\omega_l$, $\Delta_{2}=\omega_{31}+\nu_{c} - \omega_l$, $\gamma_{1}$ ($\gamma_{2}$) is the decay rate of the atomic state transition $\vert 2 \rangle \leftrightarrow \vert 1 \rangle$ ($\vert 2 \rangle \leftrightarrow \vert 3 \rangle$), and $\gamma_{m}$ is the damping rate of the oscillating mirror. Note that, in order to obtain the expectation values of the operators in the above set of equations, we drop the Hermitian Brownian noise and input vacuum noise terms which are averaged to zero. In order to obtain the steady-state solutions which are exact for the strong driving $\Omega_l$ and to the lowest order in the weak probe $\varepsilon_p$, we make the ansatz \cite{Boyd}:
\begin{equation}
\langle h \rangle = h_s + h_- e^{-i\delta t} + h_+ e^{i\delta t}, \label{cmean} 
\end{equation}
where, $h_s$ describes the steady-state solutions of any of the operators when $\varepsilon_p=0$. Moreover, $h_\pm$ are much smaller than $h_s$ and are of the same order as $\varepsilon_p$. By using the above ansatz, we obtain the following steady-state solutions
\begin{equation}\label{cs}
c_s=\dfrac{\Omega_l}{\kappa + i \Delta + \frac{g_{1}^2}{(\gamma_a+i\Delta_{1}+\frac{g_{2}^2}{\gamma_{2}+i\Delta_{2}})}},
\end{equation}\label{cm}
\begin{equation}\label{cm}
c_-=\frac{1}{d(\delta)}[(\kappa -i(\Delta +\delta) + A)(\delta ^{2}-\omega_{m}^{2} + i\delta \gamma _{m}) - 2i\omega _{m}\beta],
\end{equation}
where,
\begin{subequations}\label{B}
\begin{align}
\Delta & = \Delta_c-\dfrac{g_{o}^2}{m\hbar \omega_m^2}\vert c_s \vert^2, \\
d(\delta) &= [(\kappa -i(\Delta +\delta) + A)(\kappa + i(\Delta - \delta) + B)]\times \nonumber \\ & [(\delta^{2} - \omega
_{m}^{2} + i\delta \gamma _{m})] - 2i\omega _{m}\beta(2i\Delta - A + B) \\
A &=\dfrac{g_{1}^2 (\gamma_2-i\Delta_{2}-i\delta)}{(\gamma_1 - i\Delta_{1} - i\delta)(\gamma_2 - i\Delta_{2} - i\delta) + g_{2}^2}, \\
B &=\dfrac{g_{1}^2 (\gamma_2 + i\Delta_{2} - i\delta)}{(\gamma_1 - i\Delta_{1} + i\delta)(\gamma_2 + i\Delta_{2} - i\delta) + g_{2}^2}.
\end{align}
\end{subequations}
\begin{figure*}[t]
\includegraphics[width=0.99\textwidth]{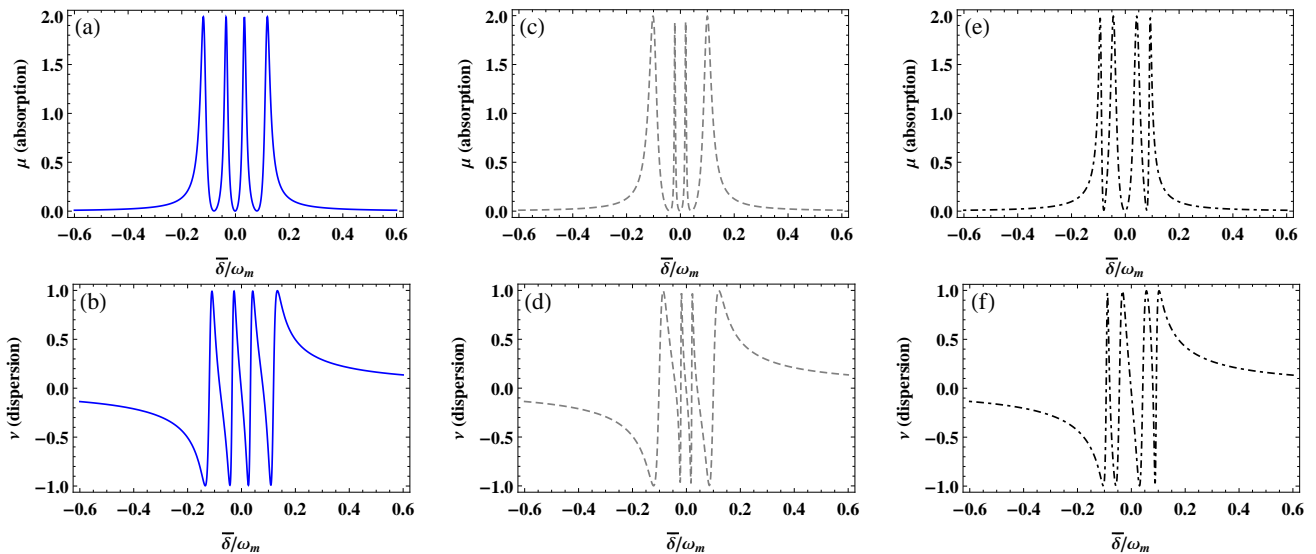}
\caption{Absorption ($\mu$) and dispersion ($\nu$) profiles are shown versus the normalized probe field detuning $\overline{\delta} / \omega_m$ displaying multiple EIT phenomenon: (a,b) For $g_{1}/2\pi = g_{2}/2\pi = 8$ MHz, we observe equally sapced multiple EIT windows. (c,d) For $g_{1}/2\pi =4$ MHz and $g_{2}/2\pi = 8$ MHz, side peaks in multiple EIT spectra broaden whereas central peaks become narrow. (e,f) For $g_{1}/2\pi =8$ MHz and $g_{2}/2\pi = 4$ MHz, side peaks in multiple EIT spectra become narrow and central peaks broaden. The corresponding experimental parameteric values are \cite{Thompson,Astafiev}: $\Omega_l/2\pi=20$ MHz, $g_{o}/2\pi=2$ MHz, $\omega_m /2\pi=100$ MHz, $\Delta_c/2\pi=100$ MHz, $\Delta_{1}=\Delta_{2}=\omega_m$, $\kappa/2\pi=4$ MHz, $\omega_{m} / \gamma_m=6700$ and $\gamma_1/2\pi=\gamma_2/2\pi=0.01$ MHz.} \label{EIT1}
\end{figure*}
The expressions (\ref{cs}) and (\ref{cm}) lead us to study the output field at the probe frequency. The response of the system to all frequencies can be detected by the output field, which can be obtained via standard input-output theory \cite{Agarwal2},
\begin{equation}
\langle c_{out}(t) \rangle  + \dfrac{\Omega_l}{\sqrt{2\kappa}} + \dfrac{\varepsilon_p}{\sqrt{2\kappa}} e^{-i\delta t}=\sqrt{2\kappa}\langle c \rangle. \label{in-out}
\end{equation}
We can express the mean value of the output field from Eqs.~ (\ref{cmean}) and (\ref{in-out}) as
\begin{eqnarray}
\sqrt{2\kappa}\langle c_{out}(t) \rangle&=&(2\kappa c_s - \Omega_l) + (2\kappa \dfrac{c_-}{\varepsilon_p} - 1)\varepsilon_p e^{-i\delta t} \nonumber \\
&+& 2\kappa (\dfrac{c_+}{\varepsilon_p^*})\varepsilon_p^* e^{i\delta t}. \label{out}
\end{eqnarray}
Note that, the second term on the right side in the above expression corresponds to the response of the whole system to the weak probe field at frequency $\omega_p$ via the detuning $\delta=\omega_p - \omega_l$. Thus, the real and imaginary parts of the amplitude of this term accounts for the absorption
and dispersion of the whole system to the probe field. We write the amplitude of the rescaled output field corresponding to the weak probe field as
\begin{equation}
\varepsilon_{out} = \dfrac{2}{\varepsilon_p} \kappa c_-. \label{eout}
\end{equation}
The corresponding real and imaginary parts of the output probe field are,
$\mathrm{Re}(\varepsilon_{out})=\frac{\kappa(c_- + c_-^*)}{\varepsilon_p}$, and $\mathrm{Im}(\varepsilon_{out})=\frac{\kappa(c_- - c_-^*)}{i\varepsilon_p}$, which correspond to absorption and dispersion respectively. These two quadratures of the output field can be measured by homodyne detections \cite{Milburn}.
\section{Multiple electromagnetically induced transparency windows}\label{sec3}
In this section, we explain multiple EIT phenomenon which emerges due to the interaction between the cavity field and mechanical mode, and due to the addition of three-level atom in the system. We write the output field at the probe frequency as
\begin{equation}
\varepsilon_{out}=\mu + i \nu. \label{Eout}
\end{equation}
Here, the real and imaginary parts, $\mu$ and $\nu$ respectively, account for the inphase and out of phase quadratures of the output field at probe frequency and correspond to the absorption and dispersion as defined above. In order to demonstrate EIT in the system, we consider the parameters from the experiments reported in \cite{Thompson,Astafiev}. The optomechanical coupling $g_{o}/2\pi=2$ MHz, $g_{1,2}/2\pi=10$ MHz, $\Omega_l/2\pi=20$ MHz, $\omega_m /2\pi=100$ MHz, $\Delta_c/2\pi=100$ MHz, $\Delta_{1,2}/2\pi=100$ MHz, $\gamma_{1,2}/2\pi=0.01$ MHz, $\kappa/2\pi=4$ MHz and $\omega_m/ \gamma_m=6700$.

In Fig.~\ref{EIT1}, we show the absorption ($\mu$) and dispersion ($\nu$) profiles as a function of the normalized detuning $\overline{\delta} / \omega_m$. The parameter $\overline{\delta} = \delta - \omega_m$ describes the probe detuning from the line centre. In the presence of optomechanical coupling ($g_{o}$), and transition coupling parameters of the TLA, $g_{1}$ and $g_{2}$, we show that three distinct (multiple) EIT windows occur in the absorption profile as shown in Fig.~\ref{EIT1}. In the presence of intense pump laser, the two transition coupling parameters $g_{1}$ and $g_{2}$ contribute together in the emergence of multiple transparency windows. For $g_1/2\pi=g_{2}/2\pi=8$ MHz, we see that the separation between the dark lines in Fig.~\ref{EIT1}(a,b) is same. However, for $g_1/2\pi=4$ MHz and $g_{2}/2\pi=8$ MHz, the central peaks become narrow whereas side peaks in the absorption profile broadens as shown in Fig.~\ref{EIT1} (c,d). In Fig.~\ref{EIT1} (e,f), we see that the central peaks broaden and side peaks become closer in the absorption and dispersion profile for $g_1/2\pi=8$ MHz and $g_{2}/2\pi=4$. Hence, in addition to the coupling parameter $g_{o}$, the power of the pump laser $P_{l}$ or $\Omega_l$ \cite{Weis,AH,Akram}, the multiple EIT windows in Fig.~\ref{EIT1}, can be controlled with transition coupling parameters $g_1$ and $g_2$ of the three-level atom.

The emergence of multiple EIT windows in hybrid optomechanics can be understood analytically as we examine the structure of the output field. As compared with the situation of single cavities \cite{Agarwal} or hybrid qubit-cavity systems \cite{Akram,Nori}, and keeping in view the structure of Eqs.~(3-5), we note that $c_s$ and $c_-$ are modified. Hence, for the case of $\Delta_{1,2}=\omega_m=\delta-x$, the numerator of the output field becomes cubic in $x$. Unlike in the case of single cavity \cite{Agarwal} and two-level atom-cavity hybrid system \cite{Akram,Nori}, there exists three roots in the present case. The third root occurs due to the presence of the TLA in the hybrid system, which has additional interfering pathway via coupling $g_{2}$. Therefore, we observe three distinct multiple EIT windows in the absorption profile, as shown in Fig.~\ref{EIT1}(a,c,e). Here each of the three roots indicates the location of the three minima for multiple EIT windows. Thus, analytically the structure of the output field confirms the splitting of usual EIT window into multiple distinct EIT windows. The three windows occur at resonances $\Delta=\omega_m$ and $\Delta_{1}=\Delta_{2}=\omega_m$, and for $g_{1}=g_{2} \neq 0$. In Fig.~\ref{EIT1}, we see that by tuning the coupling parameters $g_{1}$ and $g_{2}$, width of the multiple EIT windows can be controlled. In addition, the atom (TLA) contributes in the multiple interference, since $\Delta_{1}=\omega_m$ as well as $\Delta_{2}=\omega_m$ and $g_1 \neq 0$, $g_2 \neq 0$. Hence, the interference between the subsystems, namely; two transition coupling parameters of the atom, moving mirror, pump and probe fields respectively, corresponding to $\Delta_{1}$, $\Delta_{2}$, $\omega_m$, $\omega_l$, and $\omega_p$ causes the splitting of single EIT window, and emergence of multiple EIT windows. Therefore, dispersion in the output field changes from normal to anomalous \cite{FG}.

We also note that the presence of TLA provides a unique opportunity to switch from three to double and single EIT windows in a controllable fashion in a single experimental setup. For instance, as we switch off the transition $g_{2}$ as $g_2=0$, we observe that the resonant character of the probe field changes from three EIT windows to double EIT windows (see Fig.~\ref{trans}). This reduces the system to hybrid optomechanical system with a two-level atom as reported in Ref.~\cite{Akram,Nori}. The convergence of three windows into two windows is obvious as the structure of output field changes in a manner that its numerator and denominator goes one degree down. Hence, we can obtain the double EIT window for the case when $g_2=0$. Similarly, multiple EIT windows can be transformed to a single EIT window if transition coupling parameters $g_{1}$ and $g_{2}$ are gradually switched off. If we adjust both atomic transition couplings as $g_2=0$ and $g_1=0$, the system reduces to a single ended cavity, and a single EIT window occurs in the probe absorption. Likewise, if all the coupling parameters are gradually switched off, i.e. $g_{o}=0$ and $g_{1,2}=0$, the multi peaks in the absorption profile converge to a single peak, and a standard Lorentzian absorption peak appears with a full width at $ \overline{\delta}= \omega_m$, in agreement with a previous report \cite{Akram}. Hence, the presence of $\Lambda$-type atomic system in the optomechanical system manifests a unique opportunity to coherently control and tune the multiple EIT windows in a single experimental setup by adjusting the system parameters. This technique demonstrate great potential in promising application, for instance, multi-channel waveguides for optical signal processing, delay lines and optical buffers for telecommunication, quantum information processing and quantum metrology \cite{Fleischhauer,Huang,FG,Painter}.
\section{Slow and Fast light in hybrid optomechanics}\label{sec4}
In this section, we investigate the probe field transmission in hybrid optomechanical system with a $\Lambda$-type TLA. In the former section, we note that the second term on the right side in expression (\ref{out}) corresponds to the output field at probe frequency $\omega_p$ with detuning $\delta$. Thus, the transmission of the probe field, which is the ratio of the returned probe field from the coupling system divided by the sent probe field \cite{Painter,Weis}, is given as
\begin{equation}
t_p(\omega_p) = \frac{\varepsilon_p - \sqrt{2\kappa} c_-}{\varepsilon_p} = 1- \frac{\sqrt{2\kappa} c_-}{\varepsilon_p}.
\end{equation}
\begin{figure}[t]
\includegraphics[width=0.5\textwidth]{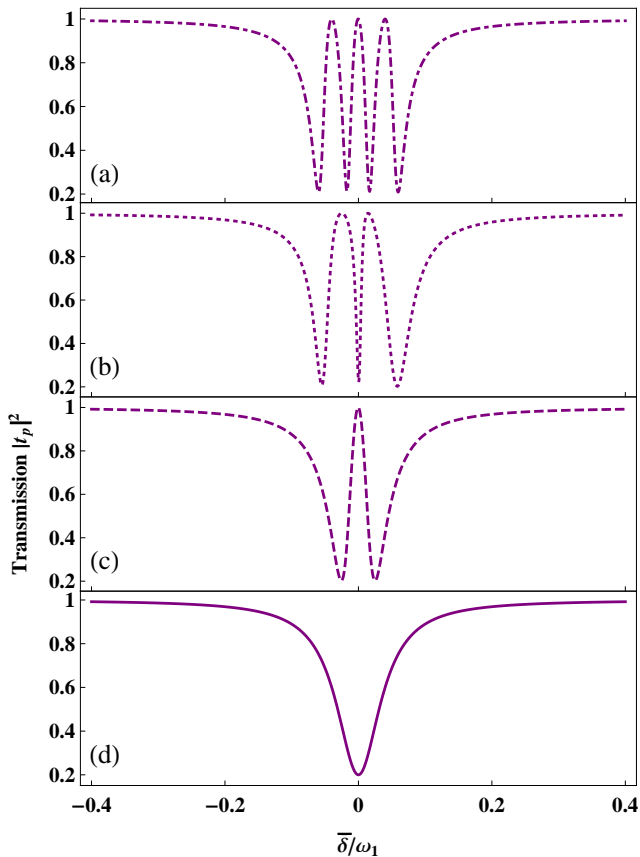} 
\caption{The transmission $\vert t_{p}\vert^2$ of the probe field as a function of normalized probe detuning $\overline{\delta}/\omega_m$ is presented: (a) For $g_{o} \neq 0$, $g_{1} \neq 0$ and $g_{2} \neq 0$, multiple (three) transparency windows occur in the probe transmission (dot-dashed curve). (b) For $g_{o} \neq 0$, $g_{1} \neq 0$ and $g_{2} =0$, double transparency windows occur (dashed curve). (c) For $g_{o} \neq 0$, $g_{1} = g_{2} = 0$, single transparency window occur (dotted curve). (d) For $g_{o} = g_{1} = g_{2} = 0$, Lorentzian curve is obtained (solid curve). All parameters are the same as in Fig.~\ref{EIT1}.}\label{trans}
\end{figure}
Moreover, for an optomechanical system, in the resonant region of the transparency window, the probe field suffers a rapid phase dispersion, viz. $\phi_t(\omega_p) = arg[t_p(\omega_p)]$. This rapid phase change $\phi_{t}(\omega_{p})$ causes the transmission group delay given as \cite{Weis},
\begin{equation} \label{phase}
\tau_g = \dfrac{d\phi_t(\omega_p)}{d\omega_p}=\dfrac{d\{arg[t_p(\omega_p)]\}}{d\omega_p}. 
\end{equation}
In the above expression, we note that the magnitude of group delay depends upon the rapid phase dispersion in the transmitted probe field. Moreover, the negative phase change indicates the anomalous dispersion which corresponds to $\tau_g < 0$ indicating the superluminal (fast light) propagation of probe field. On the other hand, the positive slope in the probe phase change expresses the normal dispersion with $\tau_g > 0$, which indicates the slow light propagation. Therefore, higher the phase dispersion, the more change in the group delay which leads to strongly altered group velocity \cite{Milonni,Akram3}.

In Fig.~\ref{trans}, we present the transmission spectrum of the probe field as a function of the normalized probe detuning $\overline{\delta}/\omega_m$. In the presence of three-level atom in the system, we see that multiple (three) narrow transparency windows appear in the probe transmission for $g_{o}/2\pi=1$ MHz, $g_{1}/2\pi=g_{2}/2\pi=4$ MHz as shown in Fig.~\ref{trans}(a). However, if we adjust the transition coupling between atomic levels $\vert 2 \rangle$ and $\vert 3 \rangle$ as $g_{2}=0$, the probe transmission displays double transparency windows as shown in Fig.~\ref{trans}(b). In a similar controllable fashion, if we adjust both atomic transition couplings as $g_{1}=0$ and $g_{2}=0$, there occur single transparency window in the probe transmission as illustrated in Fig.~\ref{trans}(c). The occurence of double and single transperacy windows is obvious as the system reduces to a hybrid system with two-level atom for $g_{2}=0,~g_{1} \neq 0$ and single ended cavity for $g_{2}=g_{1}=0$, respectively. However, if we keep optomechanical coupling as switched off ($g_{o}=0$) along with $g_{2}=g_{1}=0$, the probe transmission shows Lorentzian apsorption curve as shown in Fig.~\ref{trans}(d).
\begin{figure}[t]
\includegraphics[width=0.45\textwidth]{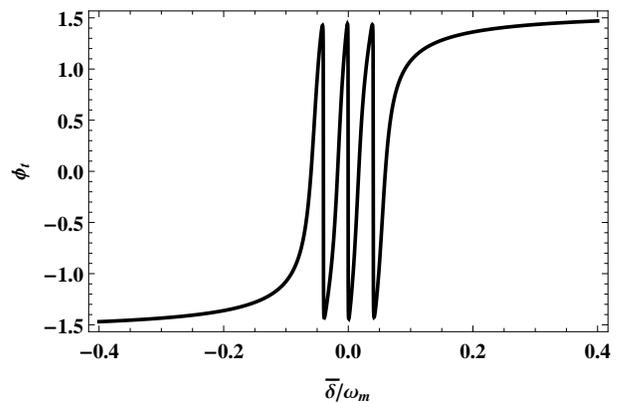}
\caption{The phase $\phi_t$ of the probe field as a function of normalized probe detuning $\overline{\delta}/\omega_m$, in the presence of three-level atom is shown. All parameters are the same as in Fig.~\ref{EIT1}.} \label{ph1}
\end{figure}

Interestingly, we see that the (transparent) multiple subluminal peaks and multiple superluminal dips simultaneously appear in the
transmission of probe field around both sides of the resonance $\overline{\delta} \sim \omega_m$. In Fig.~\ref{ph1}, we plot the phase of the transmitted probe field in the presence of three-level atom in the system for $g_{1,2} \neq 0$. We see that the phase dispersion curve in Fig.~\ref{ph1} represents anomalous dispersion with a significantly high dispersion as compare to the case of single two-level atom-cavity system \cite{Akram}. Since, in the context of fast and slow light, the high phase dispersion is advantageous as it greatly alter the group index of the medium, through which a probe pulse will propagate with strongly altered group velocity \cite{Milonni}. Therefore, we anticipate that the larger pulse advancement can be obtained in the present case.

\begin{figure}[t]
\includegraphics[width=0.45\textwidth]{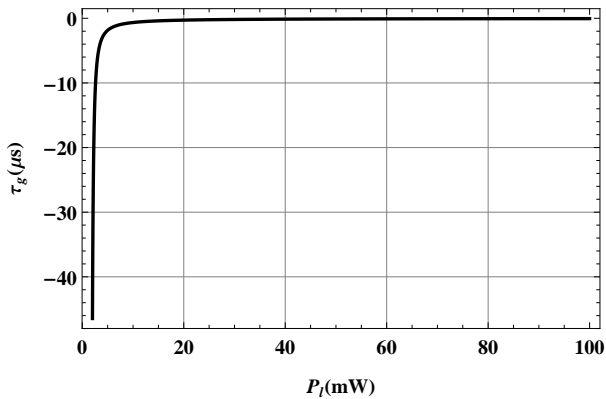}
\caption{Group delay ($\tau_g$) as a function of the pump power ($P_l$) for the case of $\Delta_{1}=\Delta_{2}=\omega_m$, indicating the superluminal propagation (pulse advancement) of the transmitted probe field. Rest of the parameters are the same as in Fig.~\ref{EIT1}.} \label{delay}
\end{figure}

In Fig.~\ref{delay}, the group delay $\tau_g$ is plotted as a function of the pump power $P_{l}$ with $\overline{\delta} = \omega_m$ and $\Delta_{1} = \Delta_{2} = \omega_m$. We see that the group delay is positive, which indicates the superluminal propagation (fast light effect) of the probe field. For the experimental realizable parameters, we obtain the pulse advancement of the order $40~\mu s$ for $g_{1}=g_{2}=4\pi \times 4$ MHz. However, previously pulse advancement of the order $20~ns$ has been reported for the case of two-level atom \cite{Akram3}. Hence, as compared to the case of single cavities and two-level atom \cite{Akram2}, the presence of TLA causes higher phase dispersion. Consequently, here we achieve significantly higher group delay as expected.
The enhanced fast light media have direct applications in real quality imaging \cite{Vogl,Glasser}, and temporal cloaking devices \cite{Fridman,Shi} that require strong manipulation of the dispersion relation, where we can envision temporally cloaking various spatial regions of an image for different durations. Similarly, by suitably adjusting the system parameters around the subluminal peaks, the probe phase turns from anomalous to normal. This reflects that the group delay changes sign from negative to positive, yielding the subluminal propagation in the probe transmission.

Thus, we infer that the superluminal behaviour of the probe field can further be enhanced by utilizing the hybridization of $\Lambda$-type atomic system in optomechanics. Hence, by employing both coupling parameters of TLA simultaneously, we achieve the high phase dispersion which leads to the higher pulse advancement than single-ended cavities and hybrid two-level atom-cavity system \cite{Akram3}.
\section{Conclusion}\label{sec5}
In conclusion, we have explained the multiple EIT phenomenon and probe field transmission in hybrid optomechanics with a three-level $\Lambda$-type atomic system. We provide full analytical model to study the absorption, dispersion and transmission profiles of the probe field. It is shown that: (i) In the presence of TLA in the system, the optomechanical coupling and the two transition coupling parameters of TLA causes the three multiple EIT windows in the probe absorption spectrum. (ii) By appropriately adjusting the system parameters, we can switch from multiple to double and single EIT windows. (iii) We study the probe transmission spectrum under wide range of system parameters. The multiple transparency windows in probe transmission simultaneously indicates transparent subluminal peaks and superluminal dips, which causes the high phase dispersion. (iv) The combined effect of the optomechanical and two transition coupling parameters of TLA, give rise to the enhancement of superluminal behaviour of probe field in the hybrid system. This work demonstrate great potential in multi-channel waveguide and optical fibers for communications networks, real quality imaging and cloaking, multi-channel quantum information processing, and delay lines \& optical buffers for telecommunication \cite{AH,Ghafoor,Milonni,Glasser,Vogl,Painter}.
\section*{Acknowledgment}
The authors would like to thank Higher Education Commission (HEC) and Quaid-i-Azam University for financial support through grant \#HEC/20-1374, and QAU-URF2014.

\end{document}